\documentstyle [12pt] {article}
\title{Threefold Post Correspondence System} 
\author{ S\'andor V\'agv\"olgyi 
\\{\sl Department of Foundations of Computer Science}
\\{\sl University of Szeged}
\\{\sl \'Arp\'ad t\'er 2, 
H-6720 Szeged, Hungary}\\
e-mail: vagvolgy@inf.u-szeged.hu\\
}

   \let\D=\Delta

\newtheorem{tet}{Theorem}[section]

\newtheorem{prop}[tet]{Proposition}

\newtheorem{conjecture}[tet]{Conjecture}

\begin{document}
\date{}
\maketitle
\section{Introduction}

Post correspondence problem  is a  basic undecidable problem \cite{davwey, rozsal, wiki}.
To establish the algorithmic unsolvability of a specific problem, 
researchers  reduce it in many cases to the Post correspondence problem.

We introduce the concept of a threefold Post correspondence system (3PCS for short) and we consider it as an instance of the  threefold Post correspondence problem.
With each 3PCS, we associate three Post correspondence systems, i.e., three instances of the Post correspondence problem.
We conjecture that for each  3PCS,  the question of the threefold Post correspondence problem  or 
for some associated  Post correspondence system  the question of the  Post correspondence problem is decidable.

In Sections \ref{tag} and \ref{teny} we present an intuitive and a formal 
 descripition, respectively,  of the concepts and the conjecture.

\section{Intuitive descripition of concepts and conjecture}\label{tag}

We define a domino as an ordered triple of strings 
written in  the top, middle, and bottom third of the domino, see Figure
 \ref{elso}.
 \begin{figure}[h]
   
\hspace{ 6 cm}
$
\left(\begin{array}{c}

abba\\

-\hspace{-2mm}-\hspace{-2mm}-\\

bbb\\

-\hspace{-2mm}-\hspace{-2mm}-\\

aa

\end{array}\right)$
 \caption{A domino.}
\label{elso}
 \end{figure}
A threefold Post correspondence system (3PCS for short) is a finite set of dominoes of this kind,  see Figure \ref{masodik}. 
\begin{figure}[h]
\hspace{ 3cm}
$
\left(\begin{array}{c}

ab\\

-\hspace{-2mm}-\hspace{-2mm}-\\

a\\

-\hspace{-2mm}-\hspace{-2mm}-\\

ab

\end{array}\right)$, 
$
\left(\begin{array}{c}

abb\\

-\hspace{-2mm}-\hspace{-2mm}-\\

babb\\

-\hspace{-2mm}-\hspace{-2mm}-\\

ab

\end{array}\right)$, 
$
\left(\begin{array}{c}

b\\

-\hspace{-2mm}-\hspace{-2mm}-\\

b\\

-\hspace{-2mm}-\hspace{-2mm}-\\

bb

\end{array}\right)$, 
$
\left(\begin{array}{c}

bba\\

-\hspace{-2mm}-\hspace{-2mm}-\\

baaa\\

-\hspace{-2mm}-\hspace{-2mm}-\\

ba

\end{array}\right)$

 \caption{A 3PCS.}
\label{masodik}
 \end{figure}
A list of dominoes may include any dominoes any number of times. 
A list of dominoes yields three words: one at
the top, one  at the middle, and one at the bottom, when we read across from 
left to right. With a  3PCS,  i.e.  a finite set of dominoes, we associate  a quadruple of 
 solitaire games played with the  dominoes:  Games  Top-Middle-Bottom,  Top-Middle,  Top-Bottom, and  Middle-Bottom.

{\bf Game Top-Middle-Bottom:}
The way to win the game is to find a list of dominoes 
where the same word appears on the top thirds as on the
middle thirds and as  on the bottom thirds of the dominoes when we read across from 
left to right.

{\bf Game Top-Middle:}
The way to win the game is to find a list of dominoes 
where the same word appears 
on the top thirds as on the
middle thirds of the dominoes when we read across from 
left to right.

{\bf Game Top-Bottom:}
The way to win the game is to find a list of dominoes 
where the same word appears 
on the top thirds as  on the bottom thirds of the dominoes when we read across from 
left to right.

{\bf Game Middle-Bottom:}
The way to win the game is to find a list of dominoes 
where the same word appears 
on the middle thirds as on the
bottom thirds of the dominoes when we read across from 
left to right.

By definition, if
 we win the game Top-Middle-Bottom, then  we  win the games 
 Top-Middle,  Top-Bottom, and Middle-Bottom as well.

\begin{figure}[h]
   \hspace{4.5cm} $
\left(\begin{array}{c}

ab\\

-\hspace{-2mm}-\hspace{-2mm}-\\

a\\

-\hspace{-2mm}-\hspace{-2mm}-\\

ab

\end{array}\right)$\hspace{-3.3mm}
$
\left(\begin{array}{c}

abb\\

-\hspace{-2mm}-\hspace{-2mm}-\\

babb\\

-\hspace{-2mm}-\hspace{-2mm}-\\

ab

\end{array}\right)$\hspace{-3.3mm}
$
\left(\begin{array}{c}

b\\

-\hspace{-2mm}-\hspace{-2mm}-\\

b\\

-\hspace{-2mm}-\hspace{-2mm}-\\

bb

\end{array}\right)$

 \caption{For the  list  $1, 2, 3$ of dominoes, 
the word $ababbb$ appears 
on the top thirds and the
middle thirds and the bottom thirds  of the dominoes when we read across from 
left to right.}\label{harmadik}
 \end{figure}
\begin{figure}[h]
    \hspace{2.5cm} 
$
\left(\begin{array}{c}

ab\\

-\hspace{-2mm}-\hspace{-2mm}-\\

a\\

-\hspace{-2mm}-\hspace{-2mm}-\\

ab

\end{array}\right)$\hspace{-3.3mm}
$
\left(\begin{array}{c}

abb\\

-\hspace{-2mm}-\hspace{-2mm}-\\

babb\\

-\hspace{-2mm}-\hspace{-2mm}-\\

ab

\end{array}\right)$\hspace{-3.3mm}
$
\left(\begin{array}{c}

b\\

-\hspace{-2mm}-\hspace{-2mm}-\\

b\\

-\hspace{-2mm}-\hspace{-2mm}-\\

bb

\end{array}\right)$\hspace{-3.3mm}
$
\left(\begin{array}{c}

ab\\

-\hspace{-2mm}-\hspace{-2mm}-\\

a\\

-\hspace{-2mm}-\hspace{-2mm}-\\

ab

\end{array}\right)$\hspace{-3.3mm}
$
\left(\begin{array}{c}

abb\\

-\hspace{-2mm}-\hspace{-2mm}-\\

babb\\

-\hspace{-2mm}-\hspace{-2mm}-\\

ab

\end{array}\right)$\hspace{-3.3mm}
$
\left(\begin{array}{c}

b\\

-\hspace{-2mm}-\hspace{-2mm}-\\

b\\

-\hspace{-2mm}-\hspace{-2mm}-\\

bb

\end{array}\right)$

 \caption{For the  list  $1, 2, 3, 1, 2, 3$ of dominoes, 
the word $ababbbababbb$ appears 
on the top thirds and the
middle thirds and the bottom thirds  of the dominoes when we read across from 
left to right.}
\label{negyedik}
 \end{figure}

\begin{figure}[h]
 \hspace{6cm}   
$
\left(\begin{array}{c}

b\\

-\hspace{-2mm}-\hspace{-2mm}-\\

b\\

-\hspace{-2mm}-\hspace{-2mm}-\\

bb

\end{array}\right)$

 \caption{For the  list  $3$ of dominoes, 
the word $b$ appears 
on the top third and the
middle third of the domino.}\label{otodik}
 \end{figure}
Consider the  3PCS shown on  Figure \ref{masodik}. 
 Figure \ref{harmadik} and Figure \ref{negyedik} show how to  win the 
game Top-Middle-Bottom.
 Figure \ref{otodik} shows how to win the 
game Top-Middle in another way.

We conjecture that for each 3PCS, 
we can decide for some game in  the associated quadruple of  solitaire games 
 whether we can win it.

\section{Formal descripition of concepts and conjecture}\label{teny}
A {\em Post correspondence system}  (PCS for short) {\em over an alphabet} $\D$
is a pair
$\langle{\bf w}, {\bf z}\rangle=\langle
(w_1, \ldots, w_n),$ $ (z_1, \ldots, z_n)\rangle $, $n\geq 1$,   of lists
of words from the  alphabet $\D$.

\noindent
 {\bf Post correspondence problem} 

{\bf Instance:} a  PCS
$\langle{\bf w}, {\bf z}\rangle$.

{\bf Question:} Do exist $k\geq 1$ and 
$i_1, \ldots, i_k$, where $1\leq i_1, \ldots, i_k\leq n$,
such that $$w_{i_1}\ldots w_{i_k}=z_{i_1}\ldots z_{i_k}\; ?$$

\noindent
Here we call the  index sequence $i_1, \ldots , i_k$ a  {\em  match } of
the PCS ${\langle{\bf w}, {\bf z}\rangle}$.
\begin{prop} \label{pcp} {\em \cite{davwey, rozsal, wiki}}
The Post correspondence problem is unsolvable.
That is, there is no algorithm which takes a 
PCS $\langle{\bf w}, {\bf z}\rangle$
as input and determines whether or not 
there is a match of the  PCS ${\langle{\bf w}, {\bf z}\rangle}$.
\end{prop}

A {\em Threefold Post correspondence system}  (3PCS for short) {\em over an alphabet} $\D$
is a triple
$\langle{\bf u}, {\bf w}, {\bf z}\rangle=\langle (u_1, \ldots, u_n),$ 
$(w_1, \ldots, w_n),$ $ (z_1, \ldots, z_n)\rangle $, $n\geq 1$,   of lists
of words from the  alphabet $\D$.

\noindent
{\bf Threefold  Post correspondence problem}

{\bf Instance:} a  3PCS $\langle{\bf u}, {\bf w}, {\bf z}\rangle$.

{\bf Question:} Do exist
 $k\geq 1$ and 
$i_1, \ldots, i_k$, where $1\leq i_1, \ldots, i_k\leq n$
such that
 $$u_{i_1}\ldots u_{i_k}=w_{i_1}\ldots w_{i_k}=z_{i_1}\ldots z_{i_k}\; ?$$

\noindent
Here we call the  index sequence $i_1, \ldots , i_k$ a  {\em  match } of
the 3PCS ${\langle{\bf u}, {\bf w}, {\bf z}\rangle}$.

To a 3PCS $\langle{\bf u}, {\bf w}, {\bf z}\rangle$, 
we assign the three PCSs $\langle{\bf u}, {\bf w}\rangle$, $\langle{\bf u}, {\bf z}\rangle$, 
$\langle{\bf w}, {\bf z}\rangle$.

\begin{conjecture} \label{gpcp}
There is an algorithm which takes a 
3PCS $\langle{\bf u}, {\bf w}, {\bf z}\rangle$ 
as input and  decides

$\bullet$ the question of the Post correspondence problem 
for the  instance 
 ${\langle{\bf u}, {\bf w}\rangle}$ or  ${\langle{\bf u}, {\bf z}\rangle}$ or 
${\langle{\bf w}, {\bf z}\rangle}$, or

$\bullet$ the question of the 
threefold  Post correspondence problem for the instance
$\langle{\bf u},{ \bf w}, {\bf z}\rangle$.
\end{conjecture}


\begin{thebibliography}{99}


\bibitem{davwey} M. D. Davis, R. Sigal,  E. J. Weyuker,
 Computability, Complexity, and Languages,
(Academic Press, New York, 1994).

\bibitem{rozsal}
Post correspondence problem. G. Rozenberg, A. Salomaa (originator), 
Encyclopedia of Mathematics.

\noindent
 URL: http://www.encyclopediaofmath.org/index.php?title=

\noindent Post\_correspondence\_problem\&oldid=13231

\bibitem{wiki}
Post correspondence problem
From Wikipedia, the free encyclopedia.

\noindent
URL: http://en.wikipedia.org/wiki/Post\_correspondence\_problem




\end{thebibliography}
\end{document}